# Outstanding Radiation Resistance of Tungsten-based High Entropy Alloys

O. El-Atwani [1*], N. Li[1], M. Li[2], A. Devaraj[3], M. Schneider[1], D. Sobieraj[4], J.S. Wrobel[4], D. D. Nguyen-Manh[5], S. A. Maloy[1], and E. Martinez[1*]

[1] Materials Science and Technology Division, Los Alamos National Laboratory, Los Alamos, NM, USA.

[2] Division of Nuclear Engineering, Argonne National Laboratory, Argonne, IL, USA

[3] Physical and Computational Sciences Directorate, Pacific Northwest National Laboratory, Richland, WA, USA

[4] Faculty of Materials Science and Engineering, Warsaw University of Technology, ul. Wołoska 141, 02-507 Warsaw, Poland

[5] Department of Materials Science and Scientific Computing, CCFE, United Kingdom Atomic Energy Authority, Abingdon, OX14 3DB, UK

## Abstract

A novel W-based refractory high entropy alloy with outstanding radiation resistance has been developed. The alloy was grown as thin films showing a bimodal grain size distribution in the nanocrystalline and ultrafine regimes and a unique 4 nm lamella-like structure revealed by atom probe tomography (APT). Transmission electron microscopy (TEM) and X-ray diffraction show an underlying body-centered cubic crystalline structure with certain black spots appearing after thermal annealing at elevated temperatures. Thorough analysis based on TEM and APT correlated the black spots with second phase particles rich in Cr and V. After both *in situ* and *ex situ* irradiation, these precipitates evolve to quasi-spherical particles with no sign of irradiation-created dislocation loops even after 8 dpa at either room temperature or 1073 K. Furthermore, nanomechanical testing shows a large hardness of 14 GPa in the as-deposited samples, with a slight increase after thermal annealing and almost negligible irradiation hardening. Theoretical modeling based on *ab initio* methodologies combined with Monte Carlo techniques predicts the formation of Cr and V rich second phase particles and points at equal mobilities of point defects as the origin of the exceptional radiation tolerance. The fact that these alloys are suitable for bulk production coupled with the exceptional radiation and mechanical properties makes them ideal structural materials for applications requiring extreme conditions.

* Corresponding author: oelatwan25@gmail.com

* Corresponding author: enriquem@lanl.gov

1. **Introduction**

Key components in magnetic fusion reactors, such as the divertor or the plasma-facing materials (PFMs), are required to have stringent properties including low activation, high melting point, good thermo-mechanical properties, low sputter erosion and low tritium retention/co-deposition. They must operate at high temperature ($\geq 1000$ K) for long durations ($> 10^7$ s), without failure or extensive erosion while exposed to large plasma heat and an intense mixture of ionized and energetic neutral species of hydrogen isotopes (D, T), He ash (fluxes $> 10^{24}$ m$^{-2}$s$^{-1}$) and neutrons[1]. Tungsten (W) is the leading PFM candidate due to its high melting temperature, low erosion rates and small tritium retention. These advantages are unfortunately coupled with very low fracture toughness characterized by brittle transgranular and intergranular failure regimes, which severely restrict the useful operating temperature window and also create a range of fabrication difficulties. Furthermore, blistering at moderate temperature (<800K) by D and He[2,3] and the formation of pits, holes and bubbles by He at higher temperature (>1600K)[4] have all been observed. The formation mechanisms that underpin these phenomena are not well understood but have largely been attributed to the accumulation of diffusing D and He in extended defects. In the slightly lower temperature range 1250-1600 K, the formation of nanometer scale bubbles is observed[5] in W exposed to He plasma. At larger He ion fluence, close to ITER[6] working conditions, exposed surfaces are found to exhibit a nanostructured surface morphology[7], termed as fuzz. The increased surface area and fragility of these nanostructured surfaces raises new concerns for the use of W as a fusion reactor PFM, particularly as a source of high-Z dust that will contaminate the plasma.

Strategies such as different alloying elements (*e.g.* W-Re, W-Ti) or nanostructure engineered W are being investigated to improve the material processing and working properties[8].

Related to the second strategy, recent work shows that state-of-the-art nanocrystalline W samples exhibit significant formation of He bubbles at the grain boundaries, which leads to de-cohesion and poor mechanical properties at operational temperatures[9–11], reducing its applicability in fusion environments. Therefore, the development of new material systems is paramount in order to enable fusion as a viable energy source.

In recent years, a novel set of alloys based on several principal elements has been developed[12–21]. The configurational entropy of mixing in multicomponent alloys tends to stabilize the solid solution based on simple underlying face-centered cubic (FCC) or body-centered cubic (BCC) crystal structures. Equiatomic compositions maximize this entropic term, promoting random solutions versus intermetallic phases or phase decomposition. Some of the HEAs show superior mechanical response to traditional materials, which in general link to dislocation properties. These materials can display high hardness values, high yield strengths, large ductility, excellent fatigue resistance and good fracture toughness. W-based refractory HEAs have been recently developed in the context of high temperature applications, showing high melting temperature (above 2873 K) and superior mechanical properties at high temperatures compared to Ni-based superalloys and nano-crystalline W[22] samples.[19,23]

High-entropy alloys have been also studied under irradiation, mostly for fcc crystalline structures. Zhang et al.[24] showed how chemical complexity can lead to a variation in the thermodynamic and kinetic properties of defects that might modify the microstructure evolution under irradiation. They linked the amount of irradiation-created defects and defect properties to electron and phonon mean free paths and dissipation mechanisms that could be tuned in these alloys by varying their composition. Granberg et al.[20] combined experiments and modeling to identify the sluggish mobility of dislocation loops as the main mechanism leading to radiation tolerance in Ni-based fcc HEAs. Kumar et al.[25] showed how Ni-based fcc HEAs lead to less radiation-induced segregation and fewer voids, although hardness was observed to increase after irradiation. Other studies show results in the same direction, with HEAs improving radiation tolerance in both fcc and BCC structures.[18,26–30] Very recently, W-based quinary HEAs with diverse composition have been synthesized as potential materials for fusion applications. The authors observed the formation of Ti carbides and laves phases at large W concentrations. The authors studied the mechanical response, showing that these materials could lead to twofold improvement in the hardness and strength due to solid solution strengthening and dispersion strengthening.[31] However, refractory HEAs have never been tested under irradiation for potential uses as PFM or structural materials in nuclear fusion environments. In this work, we have developed a quaternary nanocrystalline WTaVCr alloy that we have characterized under thermal conditions and after irradiation. We show how this alloy can be synthesized using Magnetron Sputtering/E-beam Evaporation Hybrid Physical Vapor Deposition System. Both Energy Dispersive Spectroscopy (EDS) analysis, which measures chemical composition, and Atom Probe Tomography (APT) indicate W and Ta enrichment in the deposited films. The EDS mapping on both surface and cross-sectional areas, and X-Ray Diffraction (XRD) results show a single BCC phase after deposition. The samples were irradiated with 1 MeV $Kr^{+2}$ *in situ* at the Intermediate Voltage Electron Microscope (IVEM) at Argonne National Laboratory up to 8 dpa (displacements per atom) with no sign of irradiation-created defects. Moreover, nanoindentation tests were also performed, showing hardness of the film on the order of ~15 GPa.

2. Results

2.1. HEA morphology and thermal stability

A detailed characterization of the morphology and phase details of the as-deposited HEA films results in a bimodal grain size distribution with ~70% of the grains with size in the nanocrystalline regime (≤ 100 nm) and some regions of ultrafine grain sizes (100-500 nm) with an underlying single BCC phase with a lattice constant of ~3.2 Å (see Supplemental Material).

Prior to irradiation, EDS (Figure 1) and APT (Figure 2) were performed to investigate the composition of this alloy. EDS line scan (Figure 1a) was performed to determine the composition in the alloy (Figure 1b), while EDS mapping (Figure 1c) shows uniform elemental composition. APT confirmed the EDS results, showing a film composition of 38% (± 0.09) W, 36% (± 0.09) Ta, 15% (± 0.05) Cr and 11% (± 0.05) V. The 3D distribution of elements, determined via APT, in the film prior to irradiation is shown in Figure 2 (a-d), while the 2D compositional maps using a 25x1x20 nm sized slice of APT data is shown in Figure 2 (e-j) where the color scale bars below each figure highlight the high and low concentration values for each element. The morphology is composed of very distinct compositional striations (layering) within the grains of ~4 nm thickness which was not observed in EDS. Evidence for element segregations to the grain boundaries are also found as shown in the three distinct grain boundaries captured by APT (Figure 2(i-l)).

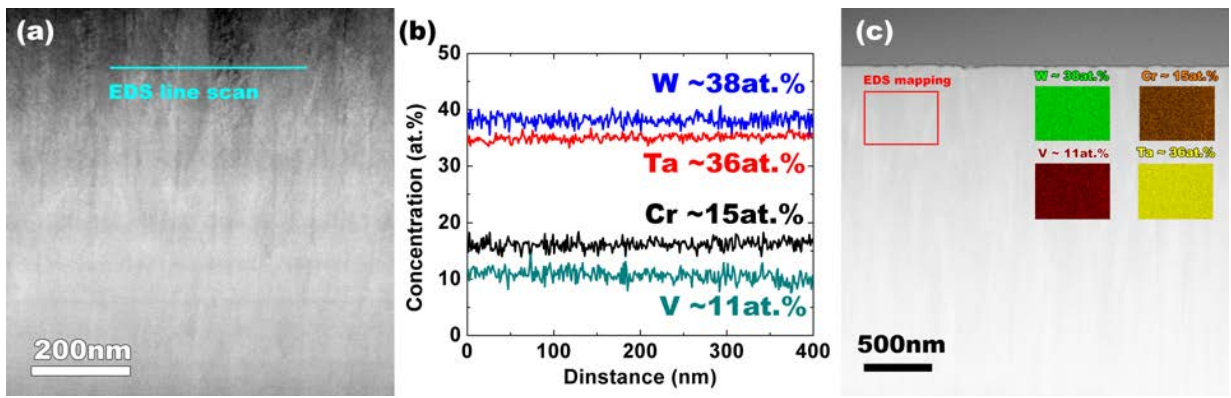

**Figure 1:** (a) Cross section TEM image of the HEA film showing a region where EDS line scan is performed. (b) EDS line scan concentration profiles of the elements in the HEA film. (c) Cross-

section scanning electron microscopy micrograph with EDS maps of the elemental composition on the HEA film.

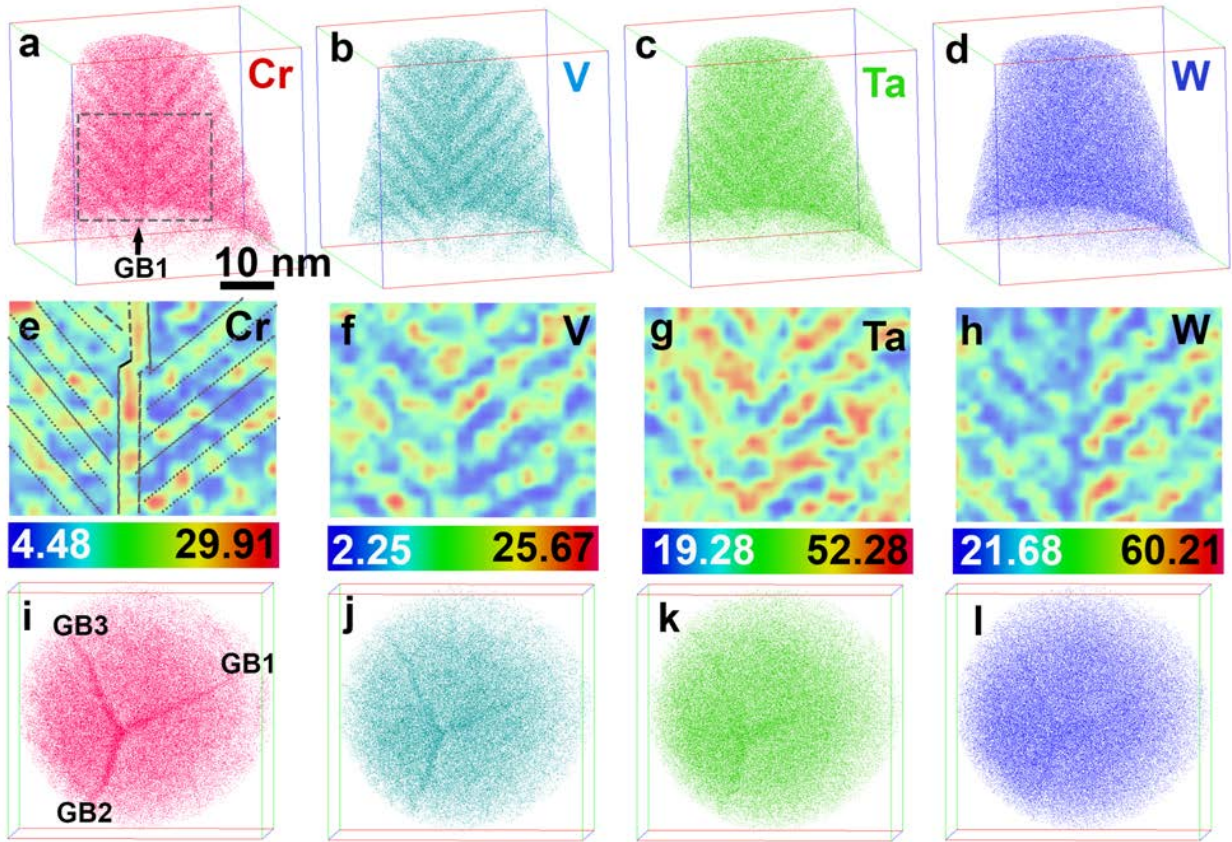

**Figure 2:** The 3D distribution of Cr, V, Ta and W in the as-deposited HEA alloy revealed by APT is shown in (a) to (d), respectively. 2D compositional maps of Cr, V, Ta and W using a 25x1x20nm slice of APT data are shown in (e) to (h) where the color scale bars below each figure denotes concentration range for each element. The top down view of the APT result showing the location of three distinct grain boundaries captured by APT as well as corresponding elemental segregation is shown in (i-l).

The thermal stability of these films was studied *in situ* in the TEM with temperatures up to 1073 K (see Figure S2 in Supplemental Material). Changes in grain sizes were not observed. Above 1023 K, some grains exhibited black spots (slight segregation of certain elements). The nanoscale distribution of elements was also checked with APT analysis for an annealed sample at 1050 K (see Figure S3 in Supplemental Material). Clear compositional layering is visible in both the ion maps and 2D composition maps. Segregation of elements to grain boundaries were also observed similar to as-deposited films. No evidence for disappearance of compositional striations or clustering of elements was observed in the APT results, indicating that compositional clustering was not homogenously distributed in the sample after heating. Not all grains demonstrated compositional clustering indicating grain variations regarding elemental segregation.

## 2.2. Irradiation

The HEA films were then irradiated *in situ* in the IVEM with 1 MeV $Kr^{+2}$ and 1073 K, with a dpa rate of 0.0006 $dpa.s^{-1}$ to 1.6 dpa (Figure S4 in the Supplemental Material). During irradiation, no dislocation loops were observed. However, further and enhanced precipitation (black spots formation) was recognized. Higher dpa irradiation was then performed on a different film at the same temperature (1073 K) with a dpa rate of 0.0016 $dpa.s^{-1}$ to 8 dpa. The morphology during irradiation is shown in Figure 3 (see also videos S1 at low dose and S2 at high dose in the Supplemental Material). Precipitation (black spots formation) was shown to occur during irradiation, which intensity increased with dose. Strikingly, dislocation loops were not observed even after 8 dpa.

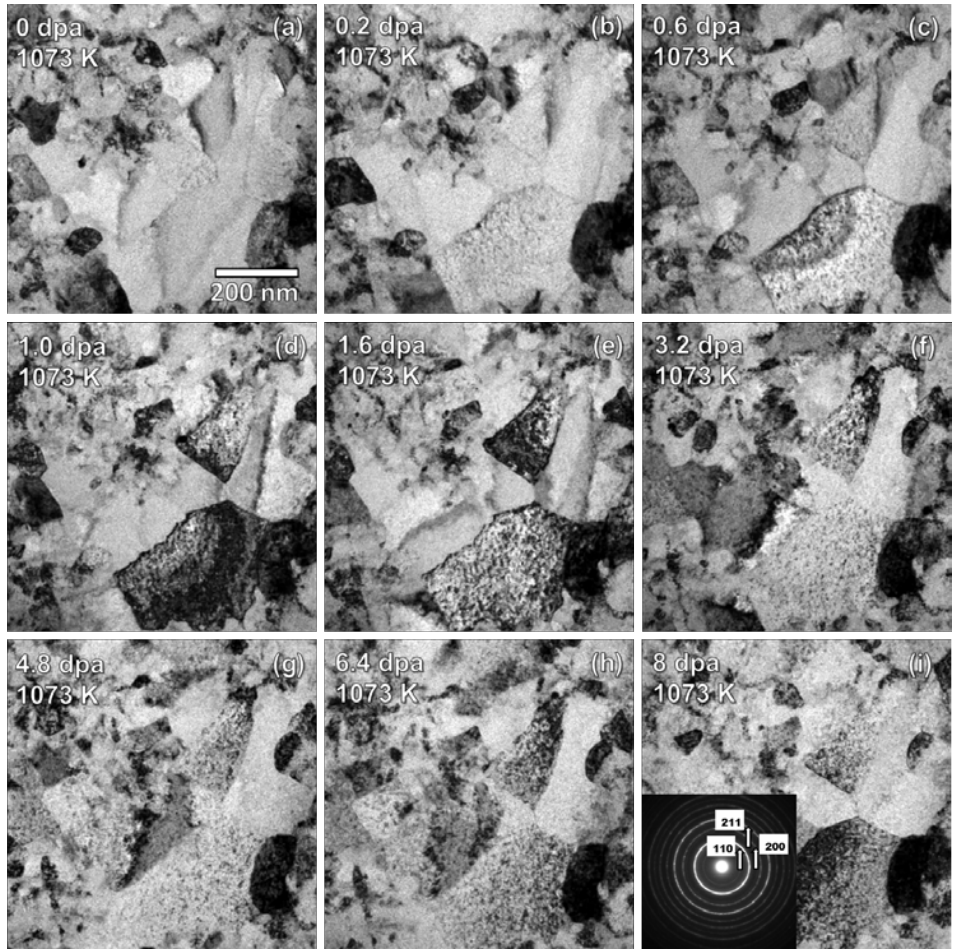

**Figure 3:** Bright-field TEM micrographs as a function of dpa of *in situ* 1 MeV Kr$^{+2}$ irradiated HEA alloy at 1073 K using a dpa rate of 0.0016 dpa.s$^{-1}$. (a) pre-irradiation, (b) 0.2 dpa, (c) 0.6 dpa, (d) 1.0 dpa, (e) 1.6 dpa and (f) 3.2 dpa, (g) 4.8 dpa, (h), 6.4 dpa and (i) 8 dpa. Images show enhanced precipitation (black spots formation) in some grains. All images have the same scale bar.

Irradiation at room temperature (RT) and dose rate of 0.0006 dpa.s$^{-1}$ to 1.6 dpa was also performed (see Figure S4 and video S3 at low dose in the Supplemental Material). No dark spots formation was observed and no dislocation loops were shown to form.

Furthermore, the mechanical properties of this material were investigated on thicker films (~3 μm) before irradiation, after annealing and after *ex situ* irradiation with 3 MeV Cu$^+$ to ~17 peak dpa (with a dose rate of 0.02 dpa.s$^{-1}$) via nanoDMA. Representative load versus displacement curves show a shift to the left in the loading curves indicating an increase in hardness, which is confirmed by the hardness versus displacement curves. The annealing process results in a hardness increase, which is slightly enhanced by irradiation (see Figure S6 in Supplemental Material).

3. Discussion

3.1. Precipitation versus loop formation

As it has been already mentioned above, the nucleation and growth of black spots was observed in the sample during annealing at elevated temperatures (1023 K). Their intensity slightly increased with irradiation and there was no apparent loop formation. Precipitates can be distinguished from dislocation loops using a TEM procedure which was detailed by Jenkins.[32] The procedure consists in analyzing the change in the **l** vector (a vector that runs from the center of the black spot to the white area) and the change in diffraction condition (**g** vector). For precipitates with symmetrical strain field, the **l** vector would follow the **g** vector when tilting the specimen to different **g** beams (vectors). However, dislocation loop **l** vectors are tied to the Burgers vector and would not rotate from one **g** vector to another. However, this procedure breaks down when precipitates have asymmetric strain fields or dislocation loops are not of edge character.[32] Moreover, in nanocrystalline samples with small grains, performing this procedure is extremely challenging. We have used two different techniques to confirm that the black spots observed in our irradiated samples are precipitates, TEM and APT analysis.

3.1.1. TEM investigation of the black spots

The HEA irradiated films are of BCC type. Although precipitation occurred, electron diffraction (Figure 3(i)) showed only BCC related rings. For a BCC material, the Burgers vector of irradiation-created dislocation loops can be of <111> or <100> type[33,34]. Therefore, there are seven possibilities of Burgers vector variants (4 for <111> type and 3 for <100> type). In W related materials, <111> type Burgers vector have been observed at 1073 K.[33,35,36] Under any diffraction condition, using the **g.b** invisibility criteria, at least 50 % of the <111> loops should be observed in the TEM image. However, in our samples, several grains showed no black spots (e.g. Figure 3).

This observation can be confirmed via 2-beam imaging in the TEM. Figure 4(a) shows a down-zone imaging of a 200 nm grain with black spots. The sample was then tilted to get a 2-beam image with the <211> **g** vector. Using <211> **g** vector, every <111> and <100> dislocation loops should be visible in the image. However, no loop was observed (Figure 4b).

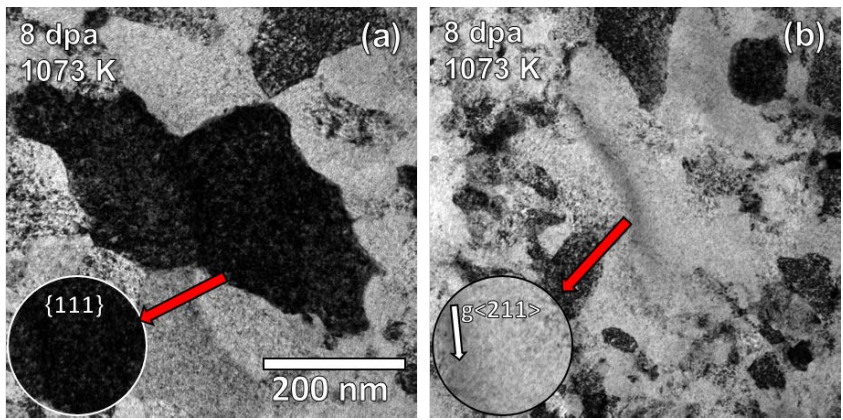

**Figure 4:** Post-irradiation bright-field TEM micrographs of 8 dpa 1 MeV Kr$^{+2}$ irradiated HEA alloy at 1073 K using a dpa rate of 0.0016 dpa.s$^{-1}$ (a) Using down <111> zone imaging showing small black spots (precipitates). (b) 2-beam image with <211> **g** vector showing no black spots. Insets show magnified images. Both images have the same scale bar.

### 3.1.2. APT analysis

APT was performed on a sample irradiated at 1050 K to ~8 dpa using 3 MeV Cu$^+$. After irradiation, no compositional layering was observed of any element (Figure 5 (a-h)). Although elemental segregation at grain boundaries is still observed, precipitation of Cr-rich phases in the grain matrices occurred as shown in the top down view of an APT reconstruction with 25 at% Cr isocomposition surface (Figure 5(m),(n)). We have also analyzed the compositional partitioning between the precipitate and the grain matrix (Figure 5(o)), showing enrichment of Cr and V and depletion of W and Ta inside the black spots. These precipitates are of ~3-5 nm size and show a density of ~ 0.03 nm$^{-2}$, both of which are consistent with the *in situ* irradiated 8 dpa sample (similar size and density). Therefore, we can conclude that the black spots observed in the irradiated HEA samples at high temperature are Cr-V rich precipitates and not irradiation-created dislocation loops.

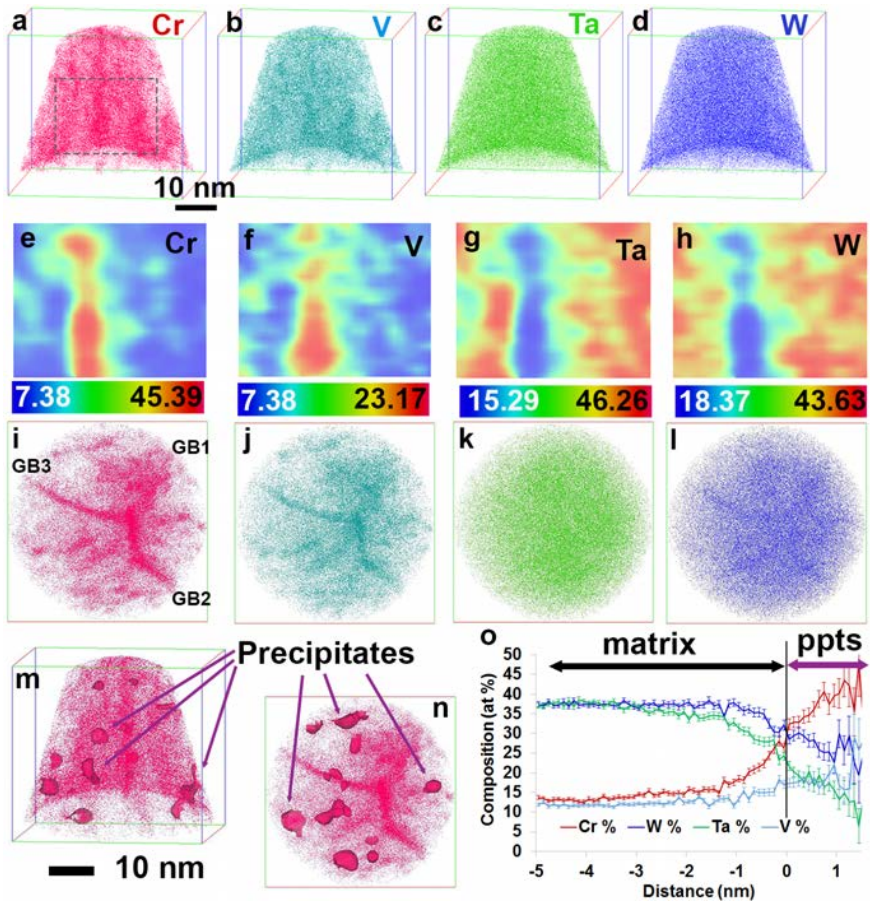

**Figure 5:** The 3D distribution of Cr, V, Ta and W in the 8 dpa irradiated HEA alloy with 3 MeV Cu$^+$ at 1050 K revealed by APT is shown in (a) to (d) respectively. 2D compositional maps of Cr, V, Ta and W using a 25x1x20nm slice of APT data is shown in (e) to (h) where the color scale bars below each figure denotes concentration values for each element. The top down view of the APT result showing the location of three distinct grain boundaries captured by APT as well as corresponding elemental segregation is shown in (i-l). (m) shows the side view of reconstruction with 25 at % Cr isocomposition surface showing Cr-V rich precipitates inside grains and the top down view is shown in (n). The compositional partitioning between the precipitate and matrix is shown in (o).

### 3.2 Origin of Cr and V rich precipitates in the HEAs alloys

To understand the origin of the Cr and V rich precipitates observed in annealed and irradiated HEAs samples, first-principles calculations of phase stability and chemical short-range orders of the multi-component W38at.%-Ta36at.%-Cr15at%-V-11at.% alloy as a function of temperature, have been systematically carried out. The cluster-expansion (CE) methodology (see Methods section and Refs.[37–41]) has been used to build up an *ab initio* based Hamiltonian with many-body effective cluster interactions (ECI) from which the configurational entropy and therefore the free energy of multi-component system can be obtained from thermodynamic integration techniques[37] in combination with semi grand canonical Monte Carlo (SGMC) simulations.

Enthalpies of mixing were calculated using the DFT and Cluster Expansion methods for nearly 270 structures within the BCC underlying crystalline lattice[42]. Values for all binary structures in the database were analyzed in order to determine the nature of the atomic interactions in all possible binary configurations (Figure S7 in the Supplemental Material). Positive and negative enthalpy of mixing values indicate a tendency for segregation or ordering, respectively.

In agreement with previous DFT/CE studies for W-Ta and W-V systems[43], it is found that the enthalpies of mixing for these two binaries are negative for the whole compositional range, trend followed also by the Cr-V binary. This behavior is characteristic of alloys between the BCC transition metals of group V and VI of the periodic table of elements from first-principles based electronic structure calculations[44]. For the Cr-Ta system, besides the small negative enthalpies of mixing, again due to mixing between elements of the two groups, there are also positive values. It is explained by the fact that along with the chemical bonding, the difference of atomic size between the 3d (Cr: $R_a$=0.130 nm) and 5d (Ta: $R_a$=0.143 nm) transition metals plays also an important role in the enthalpy of mixing. In contrast, it is found that enthalpies of mixing between the transition metals of the same group V (Ta-V binary) and VI (Cr-W binary) are positive within the whole concentration range. The analysis of the chemical ordering with the Warren-Cowley short-range order (SRO) parameters[45,46] results in a strong segregation tendency of Cr and V as can be seeing in Figure 6a (see also Supplemental Material).

We observe precipitation of Cr-V rich particles at both room temperature and 1000 K (see Figure 6(b)). To determine the local concentration of each element across the Cr-V rich phase, the average concentration over every 2D 12x12 Å cell has been calculated. The resulting concentration profile of each element is shown in Figure 6(c). The composition inside the Cr-V rich phase is found to be 62 at.% Cr, 30 at.% V, 5 at.% W and 3at.% Ta. Comparing with the compositional partitioning between the precipitates and matrix found experimentally (see Fig. 5(o)) it is observed that a very similar trend of strong Cr segregation along with V is well reproduced from our atomistic simulations. The slight difference in composition might be a kinetic effect, i.e., a result of the fact that irradiation modifies the steady state microstructure of the system.

The results of this modeling analysis, showing phase separation of Cr-V rich precipitates, are fully consistent with the compositional layering features observed in the APT results for the as-deposited samples (Figure 2(e)-(h)) and for the thermal stability investigations (Figure S3(e)-(h) in Supplemental Material). The fact that precipitation is also observed under irradiation is as well consistent with the modeling results that highlight a strong thermodynamic force for the system to phase separate.

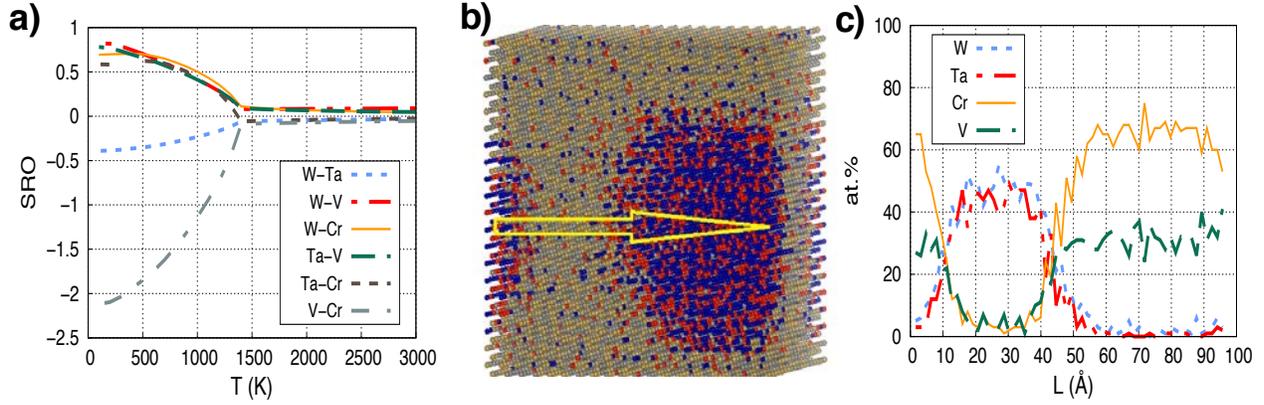

**Figure 6:** (a) Average SRO parameters in $W_{38}Ta_{35}Cr_{16}V_{11}$ alloy as function of temperature. (b) Atomic configuration in a $W_{38}Ta_{36}Cr_{15}V_{11}$ alloys at T=1000K after SGMC simulations. (c) Average concentration profile of each element along the [001] direction across the Cr-V cluster.

### 3.3 Irradiation tolerance of the HEA alloy

To correlate the experimental observations with the properties of irradiation-created defects we rely on a reaction rate model in which the reaction probability per unit time of two defects moving in 3D is given by [47]

$$R = 4\pi r_c \frac{D_i + D_j}{\Omega_a}$$

Where $r_c$ is a given capture radius, $D_i$ and $D_j$ are the diffusion coefficients of two defects, different or alike, and $\Omega_a$ is the atomic volume. For two defects of the same type this gives the probability per unit time of forming a cluster and if the defects are different this is the recombination probability. The ratio between the recombination probability and the clustering probability is maximum when the mobilities of the different defects are equal. The fact that no clustering is experimentally observed indicates that in the HEA described in this work this is a plausible case, i.e., the mobilities of vacancies and self-interstitials are similar, as oppose to what happens in pure W, with a great disparity in defect mobilities[48–50]. Maximizing recombination will lead to a reduction in defect concentration and therefore a reduction in the kinetics of precipitation and growth of second phase particles. Equal mobilities must be a consequence of the rough energy landscape induced by the local lattice distortion and the disparity in chemistry.

The presence of a high density of grain boundaries will also lead to a reduction of the defect concentrations, as grain boundaries are usually preferential recombination/annihilation sites for defects. However, even nanocrystalline W with grain sizes on the order of the ones studied here, shows the formation of large irradiation-created dislocation loops that diffuse to GBs. The fact that defect clusters are not observed in the HEA leads again to the fact that the concentration of defects inside the grains is minimized by the fact that their mobilities are similar.

It is also worth mentioning that the effect of precipitates on the recombination probability does not seem to be large since at low temperature there is no appreciable precipitate density but still there is no dislocation loop observed.

4. **Conclusions**

The present work shows the development of a new class of refractory high entropy alloy based on four elements: 38% W, 36% Ta, 15% Cr and 11% V in atomic percent measured by both Energy Dispersive Spectroscopy and Atom Probe Tomography (APT). Samples were grown using magnetron deposition from pure metal targets and characterized before and after irradiation. Microstructures present a single phase BCC crystalline structure. The films show a bimodal grain size distribution with ~70% of the grains with size in the nanocrystalline regime ($\leq 100$ nm) and some regions of ultrafine grain sizes (100-500 nm). Concurrent Transmission Electron Microscopy (TEM) and APT analysis demonstrates the existence of a second phase rich in Cr and V, first forming lamella-like regions to transform to quasi-spherical precipitates after irradiation. Irradiations were carried out both *in situ* and *ex situ* at room temperature and 1073 K up to 8 dpa.

A thorough analysis of the microstructure shows no sign of radiation-induced dislocation loops, confirming that the observed black spots are indeed second phase particles. APT results show segregation of Cr and V to grain boundaries and triple junctions. Mechanical properties of the system have been also investigated through nanoindentation. A hardness of about 14 GPa was obtained for the as-deposited sample, increasing slightly after thermal annealing and after irradiation, with small reported irradiation hardening. Accompanying modeling has also been performed. An energetic model for the four-component system has been developed based on a cluster expansion formalism. Monte Carlo simulations were subsequently carried out with the underlying energetics to find out the equilibrium properties of the system. Our DFT-based semi-grand canonical Monte-Carlo simulations show phase decomposition in remarkable agreement with the experimental results. A rate theory model tights the outstanding irradiation resistance properties to the defect mobilities and their recombination probability, which results optimal in these systems. The fact that these alloys are suitable for bulk production coupled with the exceptional radiation resistance makes them ideal structural materials for applications requiring extreme irradiation conditions.

**Acknowledgements**

This work was supported by the U.S. Department of Energy, Office of Nuclear Energy under DOE Idaho Operations Office Contract DE-AC07- 051D14517 as part of a Nuclear Science User Facilities experiment. Research presented in this article was also supported by the Laboratory Directed Research and Development program of Los Alamos National Laboratory under project number 20160674PRD3. We gratefully acknowledge the support of the U.S. Department of Energy through the LANL/LDRD Program and the G. T. Seaborg Institute for this work. D.S. and J.S.W. acknowledge the financial support from the Foundation of Polish Science Grant HOMING (No. Homing/2016-1/12). The HOMING programme is co-financed by the European Union under the European Regional Development Fund. The work at CCFE has been carried out within the framework of the EURO fusion Consortium and has received funding from the Euratom research and training programme 2014–2018 under Grant Agreement No. 633053 and funding from the RCUK Energy Programme [Grant No. EP/P012450/1]. D.N.M. would like to acknowledge the support from Marconi-Fusion, the High-Performance Computer at the CINECA headquarters in Bologna (Italy), for its provision of supercomputer resources and the Institute of Materials Science for supporting his visit to LANL. The APT sample preparation and analysis was conducted using the facilities at the Environmental Molecular Sciences Laboratory, a national scientific user facility sponsored by the DOE's Office of Biological and Environmental Research and located at Pacific Northwest National Laboratory (PNNL). APT work was funded by LDRD program at Pacific Northwest National Laboratory's Physical and Computational Sciences Directorate. A. D. would also like to acknowledge the PCSD seed LDRD funding for supporting this work.

**Methods**

1. **Experimental Methodology**

   **1.1. Preparation of W-Ta-Cr-V alloy**

   The films were prepared via magnetron deposition from pure metal targets of 99.99 % purity using deposition powers of 192, 312, 277 and 300 Watts for Cr, V, Ta and W respectively and a bias radio frequency (RF) power of 20 Watts. The deposition was performed at room temperature (RT) and 3 mTorr pressure. Approximately 100 nm films were prepared on NaCl salt using the above parameters and 60 seconds deposition. For the TEM work, TEM samples of the 100 nm films deposited on NaCl were prepared by floating the film on a standard molybdenum TEM grid using 1:1 ethanol/water solution. For X-ray diffraction and nanoindentation work, thicker films were deposited on silicon using the same parameters with 3000 seconds deposition time.

   **1.2. *In situ* TEM/Irradiation**

   *In situ* $Kr^{+2}$ ion irradiation with 1 MeV was performed using the intermediate voltage electron microscope (IVEM) attached to a Tandem accelerator at Argonne National Laboratory. The experiments were performed at RT and 1073 K. The electron beam energy was 300 keV. The average dose rate was 0.0006 dpa.s$^{-1}$ and 0.0016 dpa.s$^{-1}$ for the samples irradiated to a final dose of 1.6 dpa and 8 dpa, respectively. A CCD camera with 4k x 4k resolution was used to capture the video and still images at different doses. The dpa values were calculated (see Supplemental

Material Figures S10 and S11) using the Kinchin-Pease model in the Stopping Range of Ions in Matter (SRIM) Monte Carlo computer code (version 2013)[51] and 40 eV[52] was taken as the displacement threshold energy for all elements. The atomic percentages of the elements were 38% W, 36% Ta, 15% Cr and 11% V. Prior to the *in situ* experiment, the samples were annealed *in situ* inside the TEM with a Gatan Heating holder at 1123 K.

### 1.3. *Ex situ* Irradiation

*Ex situ* irradiation on the ~3 μm thick HEA films were performed in the Ion Beam Materials Laboratory (IBML) at Los Alamos National Laboratory (LANL) using the Tandem accelerator with 3 MeV $Cu^+$ ions at nominal incidence. The irradiations were performed at 773 K and 1050 K using a dose rate of 0.0167 dpa.s$^{-1}$. The dpa of the first 100 nm (to have similar dpa to the 100 nm TEM samples in the *in situ* TEM/irradiation work) was ~7.5 dpa which corresponds to a peak dose of ~17 dpa at 650 nm (see Figure S11 in the Supplemental Material).

### 1.4. Characterization

Characterization of the films prior to irradiation was performed using X-ray diffraction (XRD) and TEM to investigate the existing phases in the films and the overall microstructure. In addition to the *in situ* images taken at the IVEM facility, post characterization was also performed using FEI Tecnai F30 TEM operating at 300 keV, and FEI Titan 80-300 TEM operating at 300 keV. APT was performed on three HEA films: 1) pristine, 2) annealed to 1050 K, and 3) irradiated with 3 MeV $Cu^+$ to ~8 dpa. Needle specimens for APT analysis were prepared using an FEI Quanta 3D dual beam focused ion beam (FIB) system through a lift-out and annular milling process[53]. APT analysis was conducted using a CAMECA LEAP 4000XHR system in laser assisted mode using 40pJ laser pulse energy, with a 355nm UV laser at 125 KHz pulse frequency while maintaining the specimen temperature at 40K and detection rate at 0.005atoms/pulse. The APT data was reconstructed using IVAS 3.8 APT data analysis software.

### 1.5. Nanoindentation

Nanoindentation was performed on a Hysitron Tribo950 using the low force transducer, nanodynamic mechanical analysis (DMA), and a diamond Berkovich (three-sided pyramidal) tip. The nanoDMA technique allows for continuous measurement of modulus and hardness with depth by oscillating the tip at a prescribed frequency and displacement or load amplitude which is equivalent to many small unloads. A nanoDMA frequency of 100Hz and displacement amplitude of 2 nm were used. Tests were run to a final load of 10 mN and a total test time of 35 seconds using a constant strain rate load function. Indents were made on the 3 MeV $Cu^+$ irradiated samples (thick films). Samples were fixed to magnetic discs with super glue and magnetically held on the nanoindenter stage. The hardness was determined using the Oliver-Pharr method[54] with an area function calibrated on fused silica. The average hardness was taken at a displacement range of 100-150 nm. We estimate that this corresponds to interacting with material down to 450 nm (3 time the displacement) based on work by Hardie et al.[55].

### 2. Modeling Methodology

In the cluster expansion (CE) formalism the enthalpy of mixing of a K-component alloy system is defined in the form of Ising-like Hamiltonian as

$$\Delta H_{CE}(\vec{\sigma}) = \sum_{\omega} m_\omega J_\omega \langle \Gamma_{\omega'}(\vec{\sigma}) \rangle_\omega$$

where an atomic configuration is specified by a vector of the configurational variables $\vec{\sigma}$. The summation is performed over all distinct clusters $\omega$ under group symmetry operations of the underlying lattice. The parameters $m_\omega$ are multiplicities indicating the number of clusters equivalent to $\omega$ by symmetry, $J_\omega$ are the concentration independent effective cluster interaction parameters (ECIs) and $\langle \Gamma_{\omega'}(\vec{\sigma}) \rangle$ denotes the cluster functions defined as products of point functions of occupation variables on a specific cluster $\omega$ averaged over all the clusters $\omega'$ that are equivalent by symmetry to cluster $\omega$. In a $K$-component system, a cluster function is defined as a product of orthogonal point functions $\gamma_{j_i,K}(\sigma_i)$,

$$\Gamma_{\omega,n}^{(s)}(\vec{\sigma}) = \gamma_{j_1,K}(\sigma_1)\gamma_{j_2,K}(\sigma_2)\cdots\gamma_{j_{|\omega|},K}(\sigma_{|\omega|})$$

where the sequence $(s) = (j_1 j_2 \cdots j_{|\omega|})$ is the *decoration*[37] of the cluster by point functions. The number of possible decorations of clusters by nonzero point functions is a permutation with repetitions, $(K-1)^{|\omega|}$. The occupation variables and point functions are defined in such a way that it is possible to use the same formulae for $K$-component systems:

$$\gamma_{j,K}(\sigma_i) = \begin{cases} 1 & \text{if } j = 0 \\ -\cos\left(2\pi \left[\frac{j}{2}\right]\frac{\sigma_i}{K}\right) & \text{if } j > 0 \text{ and odd} \\ -\sin\left(2\pi \left[\frac{j}{2}\right]\frac{\sigma_i}{K}\right) & \text{if } j > 0 \text{ and even} \end{cases}$$

where $\sigma_i = 0,1,2,\cdots,(K-1)$, $j$ is the index of point functions $[j = 0,1,2,\cdots,(K-1)]$, and where $\left[\frac{j}{2}\right]$ denotes an operation where we take the integer plus one value of a non-integer number. To compute ECIs from first principles, the structure inversion method (SIM) is used. In SIM, energies are computed using DFT for a series of structures, the cluster functions are calculated for these structures and a set of linear equations is constructed, from which the unknown ECIs can be obtained through least-squares fitting. It is also important to note that although ECIs are assigned to ideal lattice sites they are fit to the energies of fully relaxed structures. The displacements off the ideal sites caused by size and chemical composition variations are thus included implicitly via relaxed total energy calculations. The accuracy of CE models is usually estimated by the cross-validation (CV) value that indicates the predictive power of the cluster expansion. It is defined as the square root mean of differences between those calculated from first principles and the energies predicted from the cluster expansion for structures that were not used in the fitting:

$$CV = \sqrt{\frac{1}{n}\sum_{i=1}^{n}(E_i - \hat{E}_{(i)})^2}$$

where $E_i$ is the energy of structure $i$ calculated using DFT and $\hat{E}_{(i)}$ is the energy of that structure predicted using CE.[56]

**References**

1. Baldwin, M. J. & Doerner, R. P. Helium induced nanoscopic morphology on tungsten under fusion relevant plasma conditions. *Nucl. Fusion* **48,** 035001 (2008).

2. Shu, W. M., Luo, G.-N. & Yamanishi, T. Mechanisms of retention and blistering in near-surface region of tungsten exposed to high flux deuterium plasmas of tens of eV. *J. Nucl. Mater.* **367–370,** 1463–1467 (2007).

3. Helium and hydrogen trapping in W and Mo single-crystals irradiated by He ions_Nagata_JNM307_2002.pdf.

4. Nishijima, D., Ye, M. Y., Ohno, N. & Takamura, S. Incident ion energy dependence of bubble formation on tungsten surface with low energy and high flux helium plasma irradiation. *J. Nucl. Mater.* **313,** 97–101 (2003).

5. Nishijima, D., Ye, M. ., Ohno, N. & Takamura, S. Formation mechanism of bubbles and holes on tungsten surface with low-energy and high-flux helium plasma irradiation in NAGDIS-II. *J. Nucl. Mater.* **329–333,** 1029–1033 (2004).

6. ITER - the way to new energy. *ITER* Available at: http://www.iter.org. (Accessed: 13th November 2015)

7. Takamura, S., Ohno, N., Nishijima, D. & Kajita, S. Formation of Nanostructured Tungsten with Arborescent Shape due to Helium Plasma Irradiation. *Plasma Fusion Res.* **1,** 051–051 (2006).

8. Rowcliffe-ORNL-Report.

9. El-Atwani, O. *et al.* Early stage damage of ultrafine-grained tungsten materials exposed to low energy helium ion irradiation. *Fusion Eng. Des.* **93,** 9–14 (2015).

10. El-Atwani, O. *et al.* Helium bubble formation in ultrafine and nanocrystalline tungsten under different extreme conditions. *J. Nucl. Mater.* **458,** 216–223 (2015).

11. El-Atwani, O. *et al.* In-situ TEM/heavy ion irradiation on ultrafine-and nanocrystalline-grained tungsten: Effect of 3MeV Si, Cu and W ions. *Mater. Charact.* **99,** 68–76 (2015).

12. Ranganathan, S. Alloyed pleasures: multimetallic cocktails. *Curr. Sci.* **85,** 1404–1406 (2003).

13. Yeh, J.-W. Recent progress in high-entropy alloys. *Ann. Chim. Sci. Matér.* **31,** 633–648 (2006).

14. *High-Entropy Alloys*. (Springer International Publishing, 2016).

15. Miracle, D. B. High-Entropy Alloys: A Current Evaluation of Founding Ideas and Core Effects and Exploring "Nonlinear Alloys". *JOM* **69,** 2130–2136 (2017).

16. Varvenne, C., Luque, A. & Curtin, W. A. Theory of strengthening in fcc high entropy alloys. *Acta Mater.* **118,** 164–176 (2016).

17. Xia, S., Gao, M. C., Yang, T., Liaw, P. K. & Zhang, Y. Phase stability and microstructures of high entropy alloys ion irradiated to high doses. *J. Nucl. Mater.* **480,** 100–108 (2016).

18. Lu, C. *et al.* Enhancing radiation tolerance by controlling defect mobility and migration pathways in multicomponent single-phase alloys. *Nat. Commun.* **7,** 13564 (2016).

19. Zou, Y., Ma, H. & Spolenak, R. Ultrastrong ductile and stable high-entropy alloys at small scales. *Nat. Commun.* **6,** 7748 (2015).

20. Granberg, F. *et al.* Mechanism of Radiation Damage Reduction in Equiatomic Multicomponent Single Phase Alloys. *Phys. Rev. Lett.* **116,** (2016).

21. Miracle, D. B. & Senkov, O. N. A critical review of high entropy alloys and related concepts. *Acta Mater.* **122,** 448–511 (2017).

22. El-Atwani, O., Hinks, J. A., Greaves, G., Allain, J. P. & Maloy, S. A. Grain size threshold for enhanced irradiation resistance in nanocrystalline and ultrafine tungsten. *Mater. Res. Lett.* 1–7 (2017). doi:10.1080/21663831.2017.1292326

23. Senkov, O. N., Wilks, G. B., Scott, J. M. & Miracle, D. B. Mechanical properties of Nb25Mo25Ta25W25 and V20Nb20Mo20Ta20W20 refractory high entropy alloys. *Intermetallics* **19,** 698–706 (2011).

24. Zhang, Y. *et al.* Influence of chemical disorder on energy dissipation and defect evolution in concentrated solid solution alloys. *Nat. Commun.* **6,** 8736 (2015).

25. Kumar, N. A. P. K., Li, C., Leonard, K. J., Bei, H. & Zinkle, S. J. Microstructural stability and mechanical behavior of FeNiMnCr high entropy alloy under ion irradiation. *Acta Mater.* **113,** 230–244 (2016).

26. Xia, S., Wang, Z., Yang, T. & Zhang, Y. Irradiation Behavior in High Entropy Alloys. *J. Iron Steel Res. Int.* **22,** 879–884 (2015).

27. Yang, T. *et al.* Precipitation behavior of Al x CoCrFeNi high entropy alloys under ion irradiation. *Sci. Rep.* **6,** (2016).

28. Koch, L. *et al.* Local segregation versus irradiation effects in high-entropy alloys: Steady-state conditions in a driven system. *J. Appl. Phys.* **122,** 105106 (2017).

29. Nagase, T., Rack, P. D., Noh, J. H. & Egami, T. In-situ TEM observation of structural changes in nano-crystalline CoCrCuFeNi multicomponent high-entropy alloy (HEA) under fast electron irradiation by high voltage electron microscopy (HVEM). *Intermetallics* **59,** 32–42 (2015).

30. Abhaya, S. *et al.* Effect of dose and post irradiation annealing in Ni implanted high entropy alloy FeCrCoNi using slow positron beam. *J. Alloys Compd.* **669,** 117–122 (2016).

31. Waseem, O. A. & Ryu, H. J. Powder Metallurgy Processing of a WxTaTiVCr High-Entropy Alloy and Its Derivative Alloys for Fusion Material Applications. *Sci. Rep.* **7,** (2017).

32. Jenkins, M. L. Characterisation of radiation-damage microstructures by TEM. *J. Nucl. Mater.* **216,** 124–156 (1994).

33. El-Atwani, O. *et al.* Loop and void damage during heavy ion irradiation on nanocrystalline and coarse grained tungsten: Microstructure, effect of dpa rate, temperature, and grain size. *Acta Mater.* **149,** 206–219 (2018).

34. Vetterick, G. A. *et al.* Achieving Radiation Tolerance through Non-Equilibrium Grain Boundary Structures. *Sci. Rep.* **7,** (2017).

35. Yi, X., Jenkins, M. L., Kirk, M. A., Zhou, Z. & Roberts, S. G. In-situ TEM studies of 150 keV W + ion irradiated W and W-alloys: Damage production and microstructural evolution. *Acta Mater.* **112,** 105–120 (2016).

36. Setyawan, W. *et al.* Displacement cascades and defects annealing in tungsten, Part I: Defect database from molecular dynamics simulations. *J. Nucl. Mater.* **462,** 329–337 (2015).

37. Wróbel, J. S., Nguyen-Manh, D., Lavrentiev, M. Y., Muzyk, M. & Dudarev, S. L. Phase stability of ternary fcc and bcc Fe-Cr-Ni alloys. *Phys. Rev. B* **91,** (2015).

38. Toda-Caraballo, I., Wróbel, J. S., Dudarev, S. L., Nguyen-Manh, D. & Rivera-Díaz-del-Castillo, P. E. J. Interatomic spacing distribution in multicomponent alloys. *Acta Mater.* **97,** 156–169 (2015).

39. Fernández-Caballero, A., Wróbel, J. S., Mummery, P. M. & Nguyen-Manh, D. Short-Range Order in High Entropy Alloys: Theoretical Formulation and Application to Mo-Nb-Ta-V-W System. *J. Phase Equilibria Diffus.* **38,** 391–403 (2017).

40. Wróbel, J. S., Nguyen-Manh, D., Kurzydłowski, K. J. & Dudarev, S. L. A first-principles model for anomalous segregation in dilute ternary tungsten-rhenium-vacancy alloys. *J. Phys. Condens. Matter* **29,** 145403 (2017).

41. Toda-Caraballo, I., Wróbel, J. S., Nguyen-Manh, D., Pérez, P. & Rivera-Díaz-del-Castillo, P. E. J. Simulation and Modeling in High Entropy Alloys. *JOM* **69,** 2137–2149 (2017).

42. Sobiejar, D., Wrobel, J. S. & Nguyen-Manh, D. *unpublished* (2018).

43. Muzyk, M., Nguyen-Manh, D., Kurzydłowski, K. J., Baluc, N. L. & Dudarev, S. L. Phase stability, point defects, and elastic properties of W-V and W-Ta alloys. *Phys. Rev. B* **84,** (2011).

44. Blum, V. & Zunger, A. Structural complexity in binary bcc ground states: The case of bcc Mo-Ta. *Phys. Rev. B* **69,** (2004).

45. Cowley, J. M. X-Ray Measurement of Order in Single Crystals of Cu$_3$Au. *J. Appl. Phys.* **21,** 24–30 (1950).

46. Cowley, J. M. An Approximate Theory of Order in Alloys. *Phys. Rev.* **77,** 669–675 (1950).

47. Waite, T. R. Theoretical Treatment of the Kinetics of Diffusion-Limited Reactions. *Phys. Rev.* **107,** 463–470 (1957).

48. Becquart, C. S. & Domain, C. Modeling Microstructure and Irradiation Effects. *Metall. Mater. Trans. A* **42,** 852–870 (2010).

49. Suzudo, T., Yamaguchi, M. & Hasegawa, A. Stability and mobility of rhenium and osmium in tungsten: first principles study. *Model. Simul. Mater. Sci. Eng.* **22,** 075006 (2014).

50. Huang, G.-Y., Juslin, N. & Wirth, B. D. First-principles study of vacancy, interstitial, noble gas atom interstitial and vacancy clusters in bcc-W. *Comput. Mater. Sci.* **123,** 121–130 (2016).

51. Ziegler, J. F., Ziegler, M. D. & Biersack, J. P. SRIM – The stopping and range of ions in matter (2010). *Nucl. Instrum. Methods Phys. Res. Sect. B Beam Interact. Mater. At.* **268,** 1818–1823 (2010).

52. Jung, P. 1.7 Production of atomic defects in metals by irradiation. in *Atomic Defects in Metals* (ed. Ullmaier, H.) **25,** 6–7 (Springer-Verlag, 1991).

53. Devaraj, A. *et al.* Three-dimensional nanoscale characterisation of materials by atom probe tomography. *Int. Mater. Rev.* **63,** 68–101 (2018).

54. Oliver, W. C. & Pharr, G. M. Nanoindentation in materials research: Past, present, and future. *MRS Bull.* **35,** 897–907 (2010).

55. Hardie, C. D., Roberts, S. G. & Bushby, A. J. Understanding the effects of ion irradiation using nanoindentation techniques. *J. Nucl. Mater.* **462,** 391–401 (2015).

56. van de Walle, A. & Ceder, G. Automating first-principles phase diagram calculations. *J. Phase Equilibria* **23,** 348–359 (2002).